%% file: OSAmeetings.tex
\documentclass[letterpaper,10pt]{article}

\usepackage{osameet2}
\usepackage{tikz,pgfplots}
\usepackage{lipsum}
\usepackage[percent]{overpic}

\usetikzlibrary{shadows,arrows,shapes,calc,positioning,decorations.pathreplacing,fit}
\usepackage{amsmath,amssymb}
\usepackage{graphicx}
\pgfplotsset{compat=1.14}

\begin{document}

\title{SWIFT: Scalable Ultra-Wideband Sub-Nanosecond Wavelength Switching for Data Centre Networks}

\author{Thomas Gerard*, Christopher Parsonson, Zacharaya Shabka, Polina Bayvel, Domani\c{c} Lavery and Georgios Zervas}
\address{Optical Networks Group, Dept. of Electronic and Electrical Engineering, University College London, London, UK, WC1E 7JE.}
\email{*uceetmh@ucl.ac.uk}

\vspace{-1.6em}
\begin{abstract}
%
We propose a time-multiplexed DS-DBR/SOA-gated system to deliver low-power fast tuning across S-/C-/L-bands. Sub-ns switching is demonstrated, supporting 122$\times$50~GHz channels over 6.05 THz using AI techniques.
\end{abstract}
\vspace{-.5em}

\vspace{-1.0em}
\input{Sections/Introduction.tex}
\vspace{-.5em}
\input{Sections/ExperimentalSetup.tex}

\vspace{-.5em}
\input{Sections/ResultsAndDiscussion.tex}




\end{document}

%% file: Sections/Introduction.tex
\section{Introduction}
\vspace{-.5em}

The most common data center network (DCN) packet length is $<$256 bytes which translates to 20~ns slots in 100G links \cite{clark2018}. Optical circuit switching (OCS) aims to transform data centre networks (DCNs) but needs to operate at packet speed and granularity \cite{benjamin2020}. 
Recent breakthroughs have brought OCS closer to reality. A hardware-based OCS scheduling algorithm has demonstrated synchronous scheduling of up to 32,768 nodes within 2.3~ns \cite{benjamin2020}. A clock phase caching method has enabled clock and data recovery in less than 625 ps, supporting 10,000 remote nodes \cite{clark2018}. Yet, energy-efficient, sub-ns, many-channel optical switching remains a challenge. 
Wideband fast tuneable lasers have demonstrated switching on ns timescales \cite{simsarian2006,gerard2020}, and as low as 500~ps but over limited bandwidths \cite{ueda2019}. Static laser diodes (LDs) gated by semiconductor optical amplifiers (SOAs) have achieved 912 ps 10-90\% rise-fall times with $\sim$2~ns settling time ($\pm5\%$ of the target value) \cite{shi2019}; however, the power consumption and device count limit the scalability of this approach. A similar method used an optical comb where each wavelength was filtered then gated by an SOA \cite{lange2020}; the power consumption and device count therefore also increase linearly with number of channels, limiting scalability (see Fig.~1(b)).

In this paper, we introduce SWIFT: a modular system with Scalable Wideband Interconnects for Fast Tuning. SWIFT combines pairs of optimised widely tuneable lasers (TLs), multiplexing
their wavelength reconfiguration on packet timescales. The lasers are gated by pairs of fast switching SOAs, resulting in wideband, sub-ns switching. The modular design of SWIFT (Fig. 1(a)) shows that just two lasers and two SOAs cover each optical transmission band. SWIFT power consumption is, therefore, practically independent of channel count; Fig.~1(b) shows that SWIFT becomes more power efficient than alternative sub-ns switching sources 
beyond 8$\times$50~GHz spaced channels. 
\begin{figure*}[!b]
\centering
\label{fig:principle}
\setlength\tabcolsep{0pt}
\renewcommand{\arraystretch}{0} 
    \begin{tabular}{cccc}
    \hspace{1mm}%
    \includegraphics[scale=0.54]{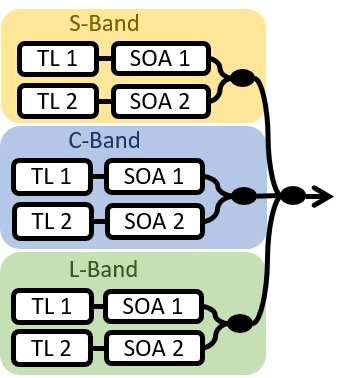}
    & 
    \hspace{0mm}%
     \includegraphics[scale=0.19]{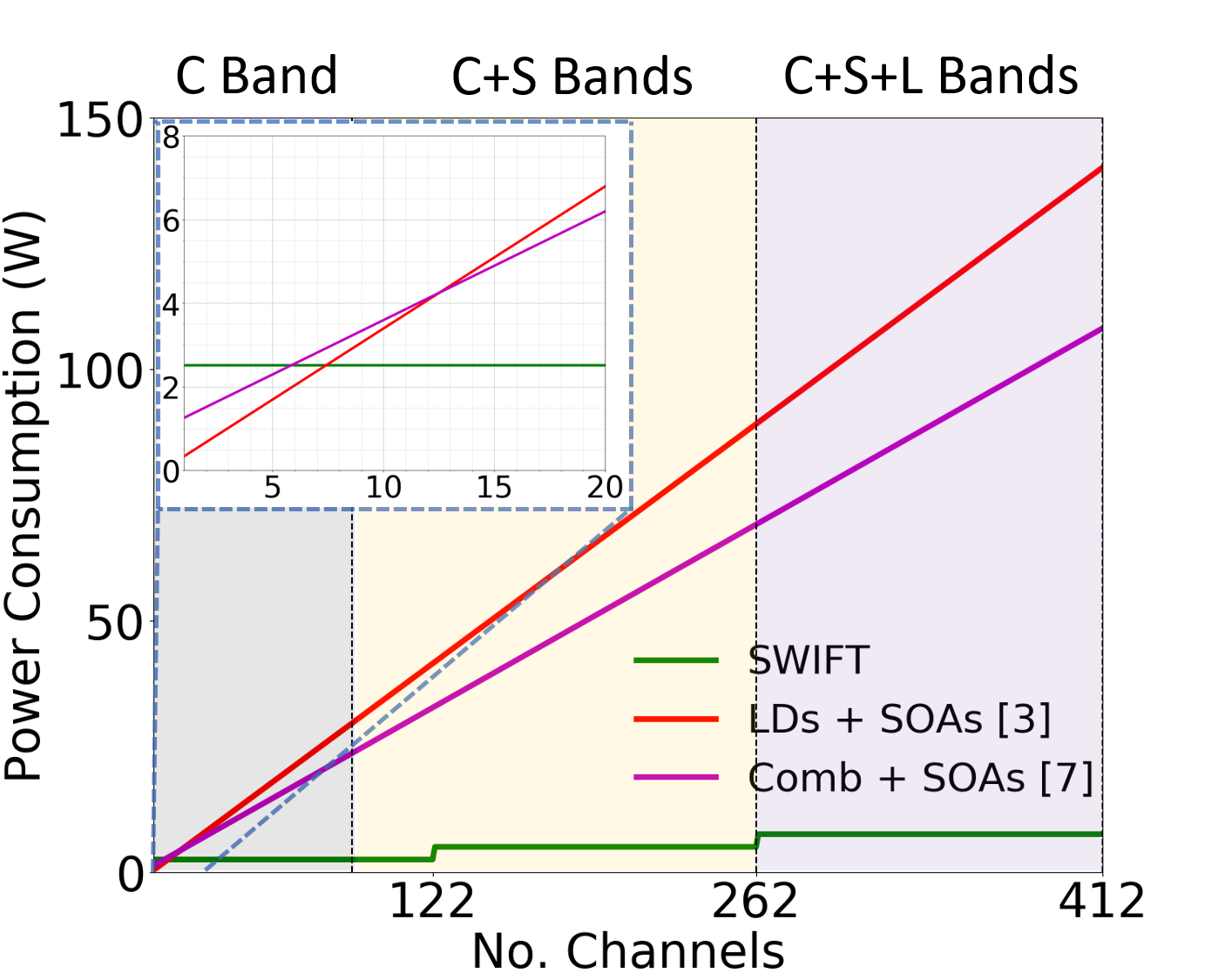}
    &
   \hspace{1mm}%
   \includegraphics[scale=0.5]{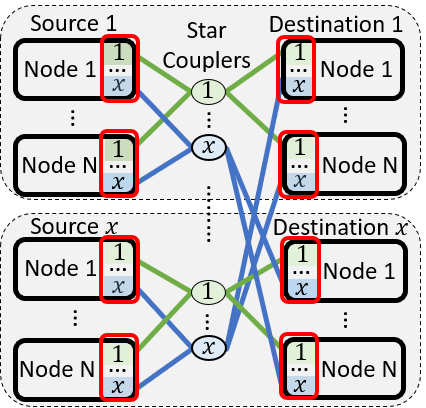}
    &
    \hspace{3mm}%
  \includegraphics[scale=0.11]{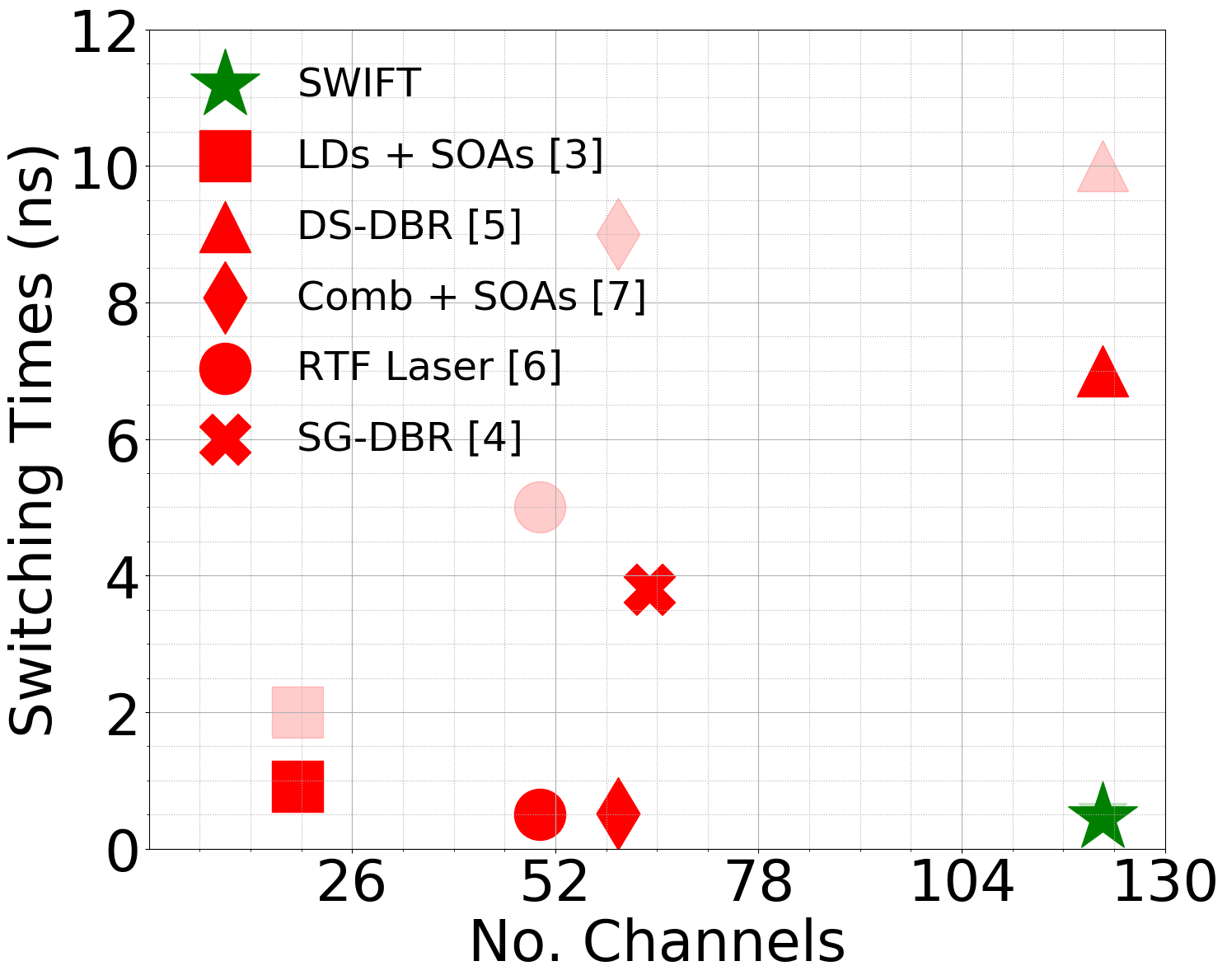}
  \\
  (a)&(b)&(c)&(d)
    \end{tabular}

\vspace{-1em}
\caption{
(a) Modular SWIFT architecture across S-, C- and L-bands.
(b) Power consumption comparison of laser switch designs vs. no. of channels, using data reported in \cite{benjamin2020arXiv}. 
(c) PULSE DCN architecture with SWIFT modules (in red). (d) Comparison of switching  times (reported rise (solid) and estimated settling  (faded)) against no of channels for different switch systems.  
}
\vspace{-2em}

 \end{figure*}

The SWIFT modules can be deployed as transmitters in DCN architectures such as PULSE \cite{benjamin2020arXiv}, as shown in Fig.~1(c). In this architecture, each node has $x$ SWIFT transmitters (highlighted in red), each local pod has $N$ nodes, and $x^2$ star couplers enable there to be $x$ source and $x$ destination pods. Thus, PULSE network's number of end-points scales with $N \times x$, where $N$ is limited by the number of wavelength channels. The large number of channels supported by SWIFT therefore allows for significant scalability in the PULSE DCN \cite{benjamin2020}.

The concept of time-multiplexed, fast tuneable lasers was proposed in \cite{benjamin2020arXiv,ryan2008}, but faced the challenge of optimising multiple lasers and SOAs for reliable fast tuning. SWIFT overcomes this by applying artificial intelligence (AI) techniques to the devices, enabling autonomous optimisation. This has allowed us to demonstrate, for the first time, a time-multiplexed, gated laser tuning system that can tune over 6.05~THz of bandwidth and consistently switch in 547~ps or better to support 20~ns timeslots. SWIFT outperforms other fast switching systems in terms of rise time, settling time and channel count, as shown in Fig~1(d). 

%% file: Sections/ExperimentalSetup.tex
\section{Experimental Setup}
\vspace{-0.5em}

The setup used to demonstrate SWIFT is shown in Fig. 2(a). A pair of commercial Oclaro (now Lumentum) digital-supermode distributed Bragg reflector (DS-DBR) lasers were driven by 250 MS/s arbitrary waveform generators (AWGs) with 125~MHz bandwidth. Detailed IV measurements were used to map supplied voltage to desired current. Each laser was connected to a commercial InPhenix SOA, supporting 69~nm of bandwidth with typical characteristics of 7~dB noise figure, 20~dB gain, and 10~dBm saturation power. Each SOA was driven with a 45~mA current source modulated by a 12~GS/s AWG with $\pm$0.5~V output and amplified to $\pm$4~V using an electrical amplifier. All four optical devices were held at 25$^\circ$C using temperature controllers. The SOAs were coupled together and passed to a digital coherent receiver (50 GS/s, 22~GHz bandwidth) and a digital sampling oscilloscope (50 GS/s, 30~GHz bandwidth), which provided optimisation feedback to the DS-DBR lasers and to the SOAs respectively.

\begin{figure*}[!t]
\centering
\label{fig:exp_results_1}
\setlength\tabcolsep{0pt}
\renewcommand{\arraystretch}{0} 
    \begin{tabular}{ccc}
    \begin{overpic}[scale=0.45]{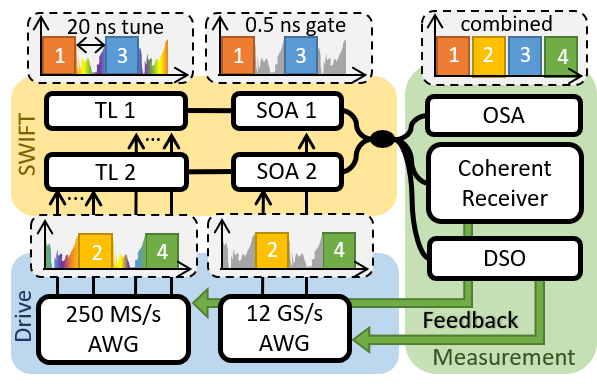}
         \put(-5,55){(a)}
    \end{overpic}
    &
     \begin{overpic}[scale=0.14]{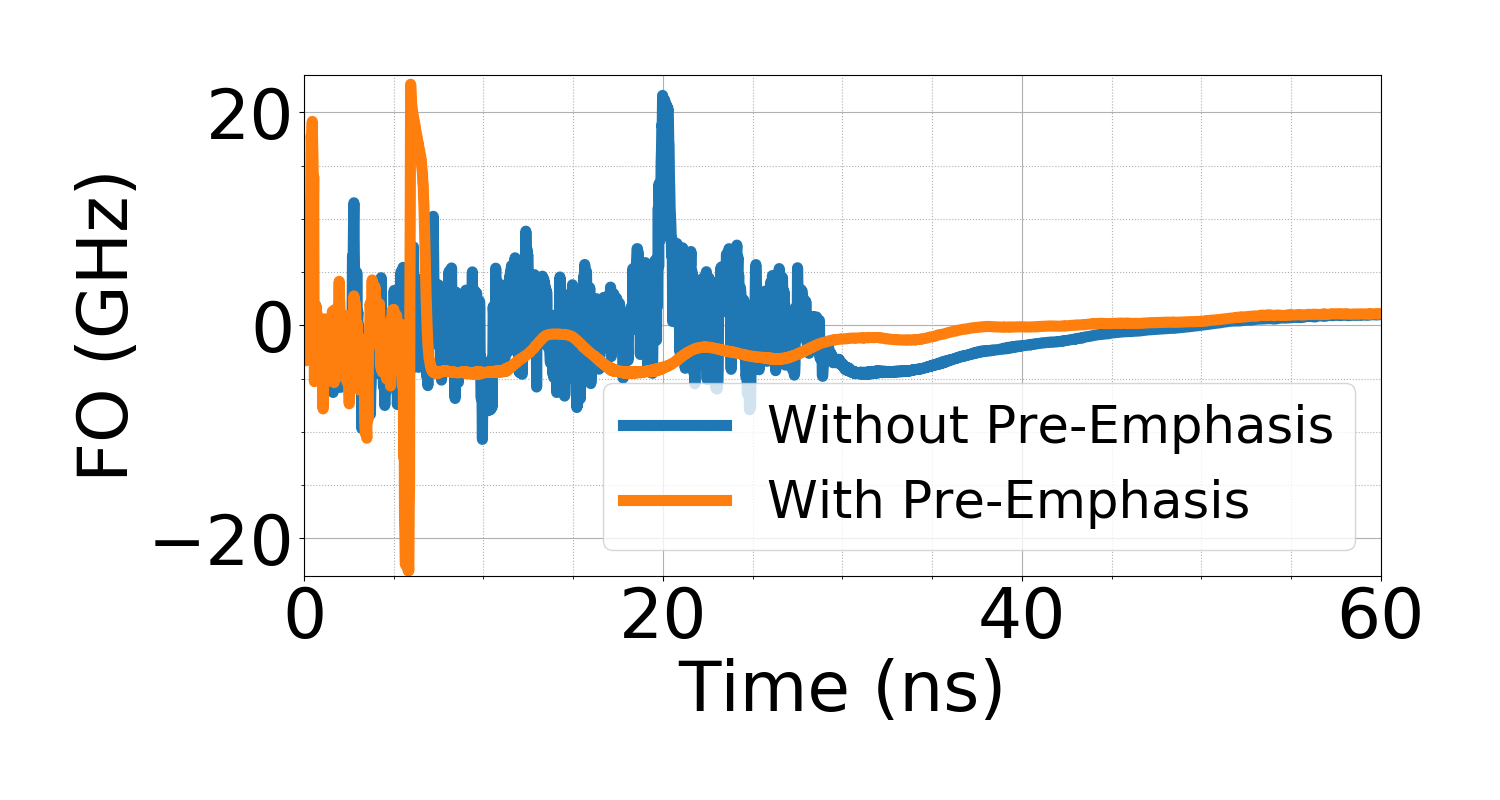}
          \put(12,50){(b)}
   \end{overpic}
    &
    \begin{overpic}[scale=0.14]{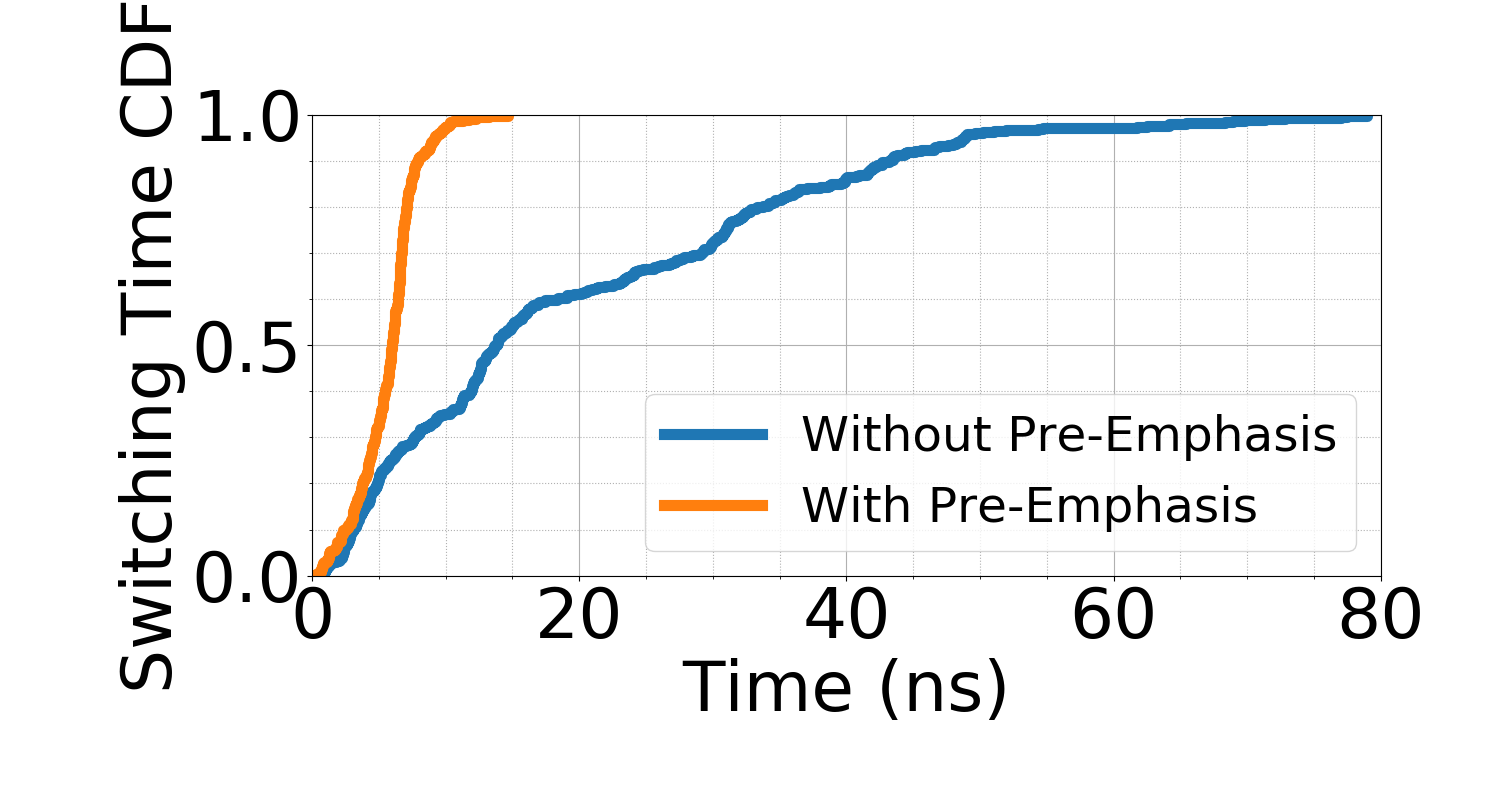}
          \put(12,50){(c)}
   \end{overpic}
  \end{tabular}
\vspace{-1em}
\caption{(a) Experimental setup of time-multiplexed SWIFT tuneable lasers (TL) gated by SOAs. (b) TL frequency offset (FO) of worst-case current swing w/ \& w/o optimiser. (c) CDF of all worst-case laser switch combinations w/ \& w/o optimiser.}
\vspace{-2em}
 \end{figure*}


%% file: Sections/ResultsAndDiscussion.tex
\section{Results and Discussion}
\vspace{-0.5em}

\subsection{Regression optimised laser switching}



Fast wavelength switching can be achieved by applying `pre-emphasis' to the drive sections of an integrated semiconductor laser. Until recently, pre-emphasis values had to be carefully tuned by hand for select samples then extrapolated \cite{simsarian2006}. 
Here, we apply a linear regression optimiser to automatically calculate the pre-emphasis values for reliable fast tuning. We measured the output of the DS-DBR laser during a switching event using the coherent receiver, then used the instantaneous frequency response as the error term within a linear regression optimiser to iteratively update the applied pre-emphasis values \cite{gerard2020}. Fig 2(b) shows an example of the laser's switching response before and after application of the optimiser.  
We applied this optimiser to 21 of the 122 supported channels, testing the extremes of lasing frequency and drive current, covering 462 any-to-any switching events across 6.05~THz (1524.11-1572.48~nm). Fig. 2(c) shows the cumulative distribution of the time taken to reach $\pm$5~GHz of the target wavelength. We measure a worst case switch time of 14.7~ns, and a worst case frequency offset after 20~ns of $-$4.5~GHz. This indicates that SWIFT is potentially suitable for burst mode coherent detection, as 28~GBd dual-polarisation quadrature phase shift keying is tolerant of frequency offsets up to $\pm$7~GHz \cite{simsarian2014}.


\vspace{-0.5em}
\subsection{Particle swarm optimised SOA switching}
\vspace{-0.3em}

SOA driving signals must also be optimised to approach their theoretical rise/fall times of $\sim100$ ps. 
Previous optimisation attempts did not consider settling times nor the ability to automate the optimisation of driving conditions for 1,000s of different SOAs in real DCNs \cite{gallep2002, figueiredo2015}. 
To solve this, PSO (a population-based metaheuristic for optimising continuous nonlinear functions by combining swarm theory with evolutionary programming) was used in this work to optimise the SOA driving signals. PSO has previously been applied to proportional-integral-derivative (PID) tuning in control theory \cite{kusuma2016}, but has not yet been used as an autonomous method for optical switch control. In the optimisation, $n = 160$ particles (driving signals) were initialised in an $m = 240$ (number of points in the signal) hyperdimensional search space and iteratively `flown' through the space by evaluating each particle's position with a fitness function $f$, defined as the mean squared error between the drive signals' corresponding optical outputs (recorded on the oscilloscope) and an ideal target `set point' (SP) with 0 overshoot, settling time and rise time. As shown in Fig. 3(a) and (b), the
$\pm$ 5\% settling time (effective switching time) of the SOA was reduced from 3.72~ns (when driven by a simple square driving signal) to 547~ps, with the 10-90\% rise time also reduced from 697~ps to 454~ps. The PSO routine required no prior knowledge of the SOA, therefore provides a flexible, automated and scalable method for optimising SOA gating.

\begin{figure*}[!t]
\centering
\label{fig:exp_results_2}
\setlength\tabcolsep{0pt}
\renewcommand{\arraystretch}{0} 
    \begin{tabular}{ccc}    
  \begin{overpic}[scale=0.15]{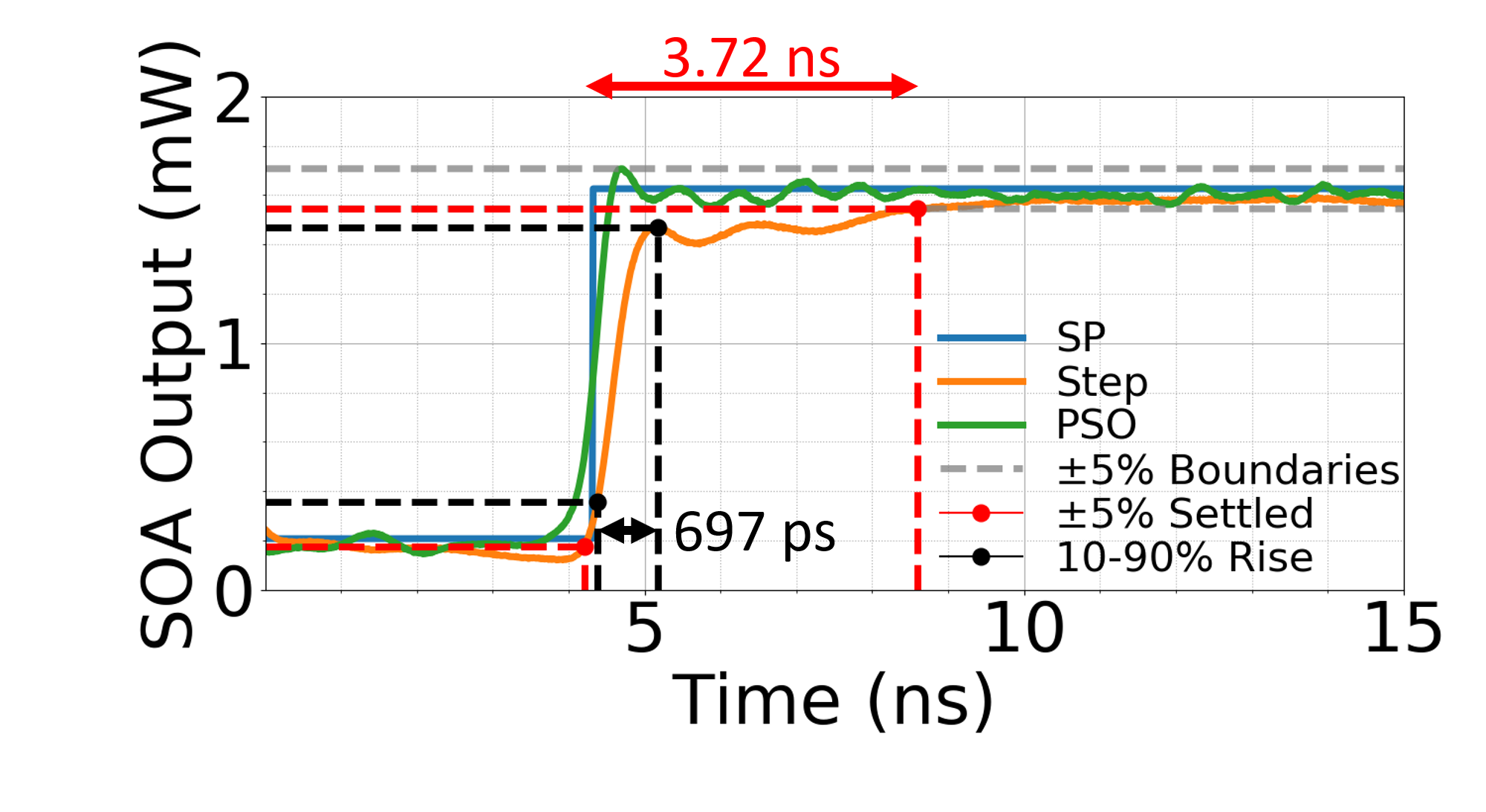}
         \put(15,52){(a)}
  \end{overpic}
    &
  \begin{overpic}[scale=0.15]{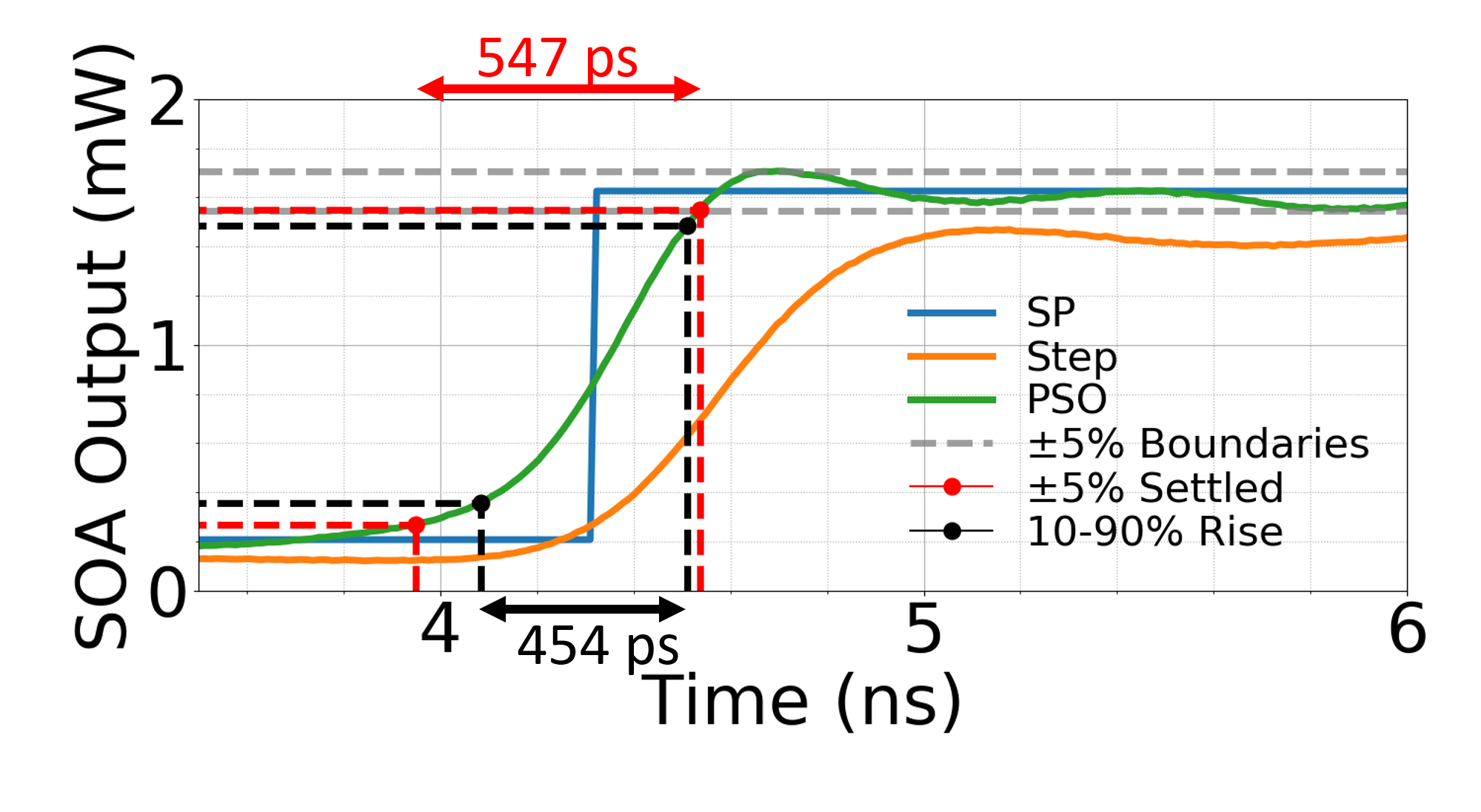}
         \put(12,52){(b)}
  \end{overpic}
    &
   \begin{overpic}[scale=0.03]{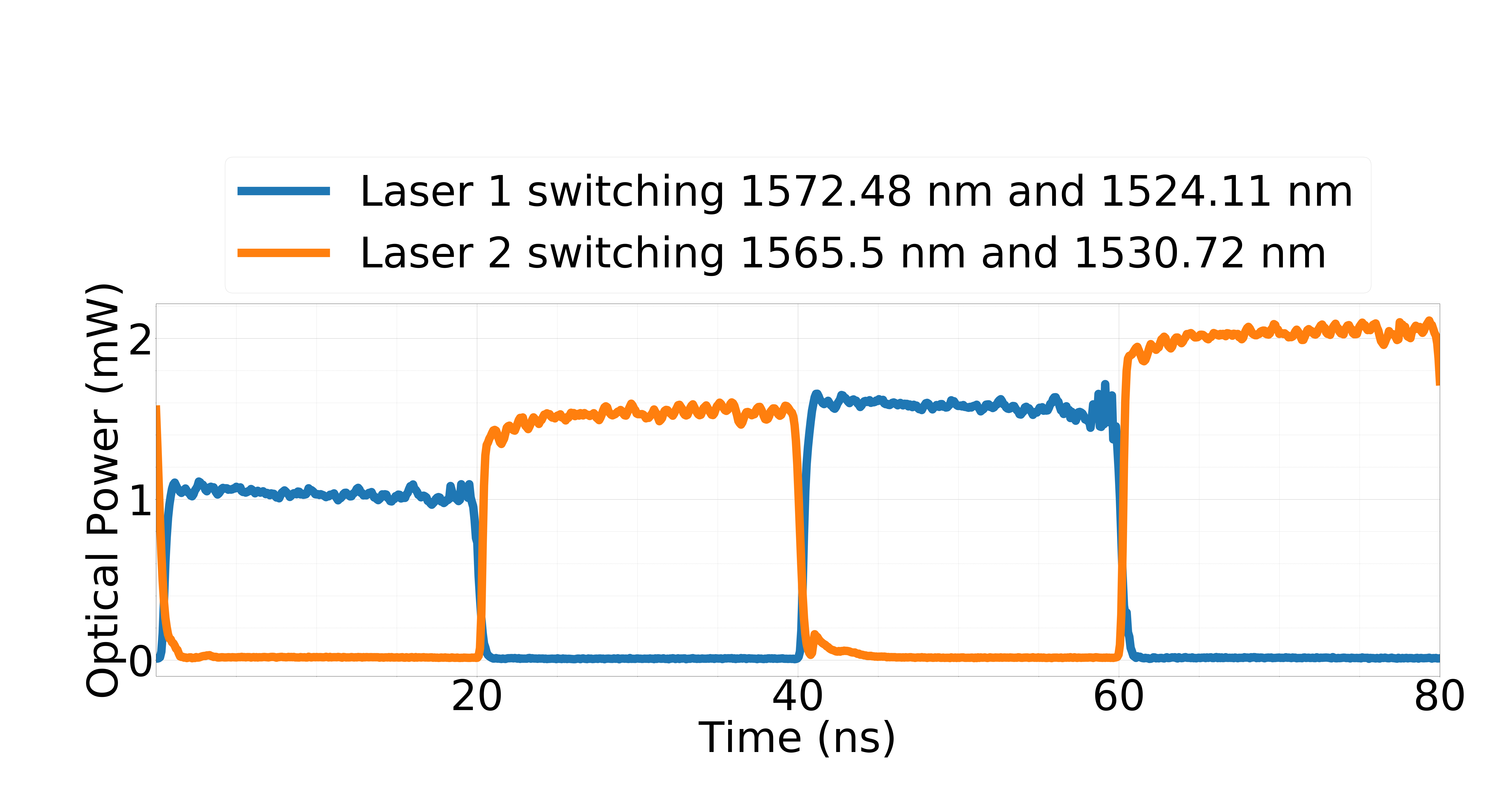}
          \put(5,40){(c)}
   \end{overpic}
  \\
     \begin{overpic}[scale=0.22]{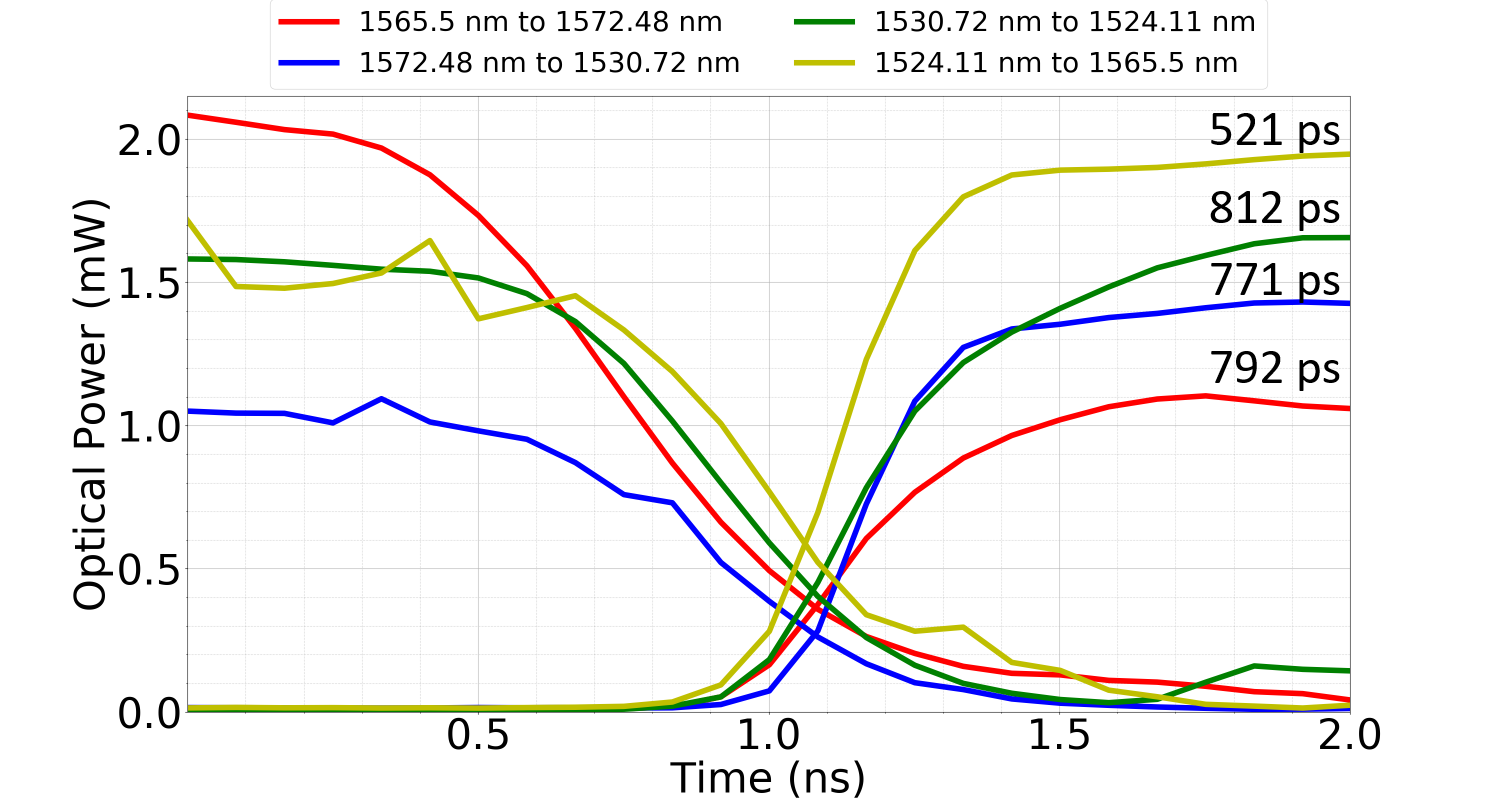}
         \put(3,45){(d)}
    \end{overpic}
    & 
  \begin{overpic}[scale=0.15]{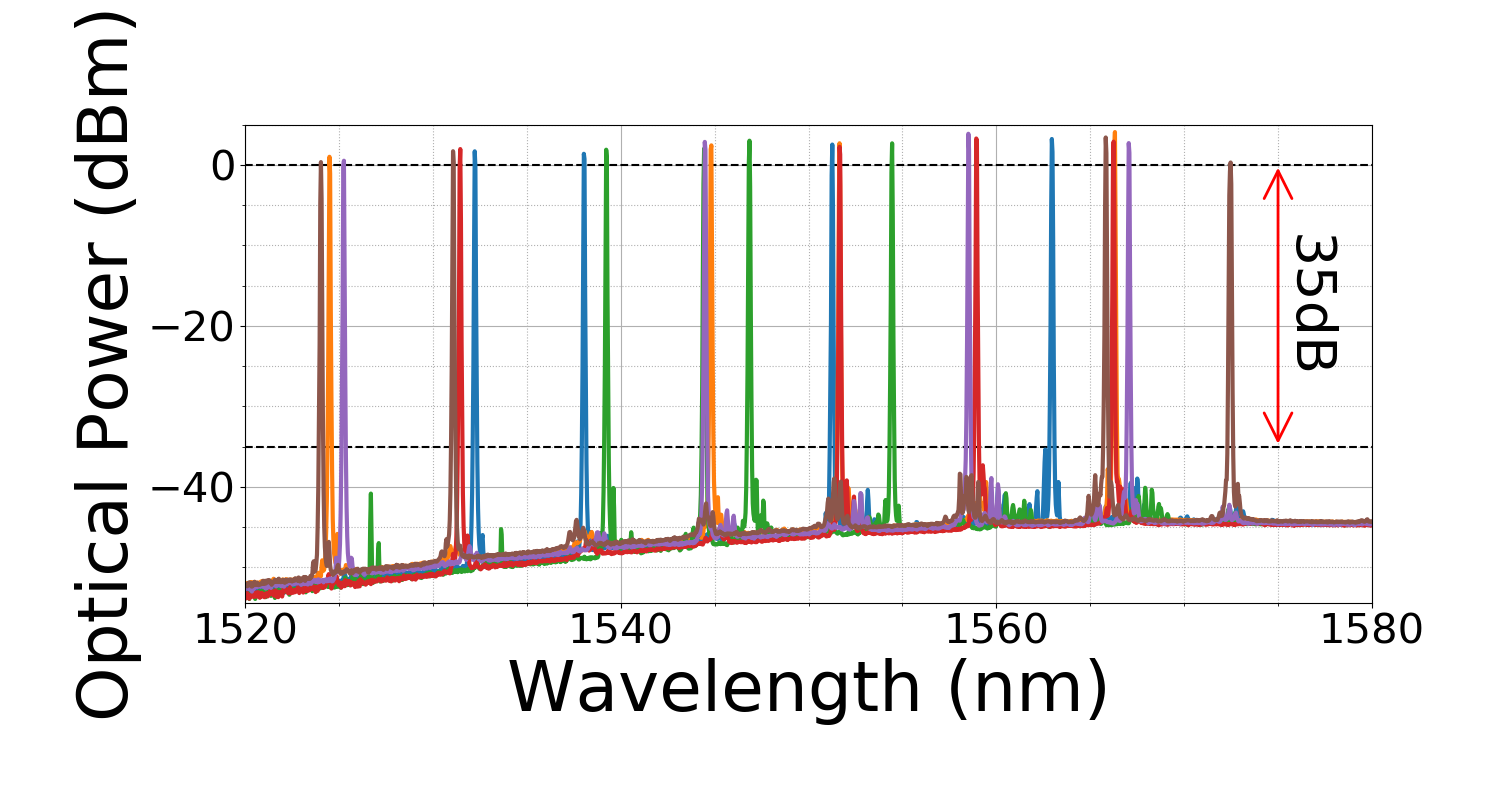}
         \put(12,50){(e)}
    \end{overpic}
    &
  \begin{overpic}[scale=0.037]{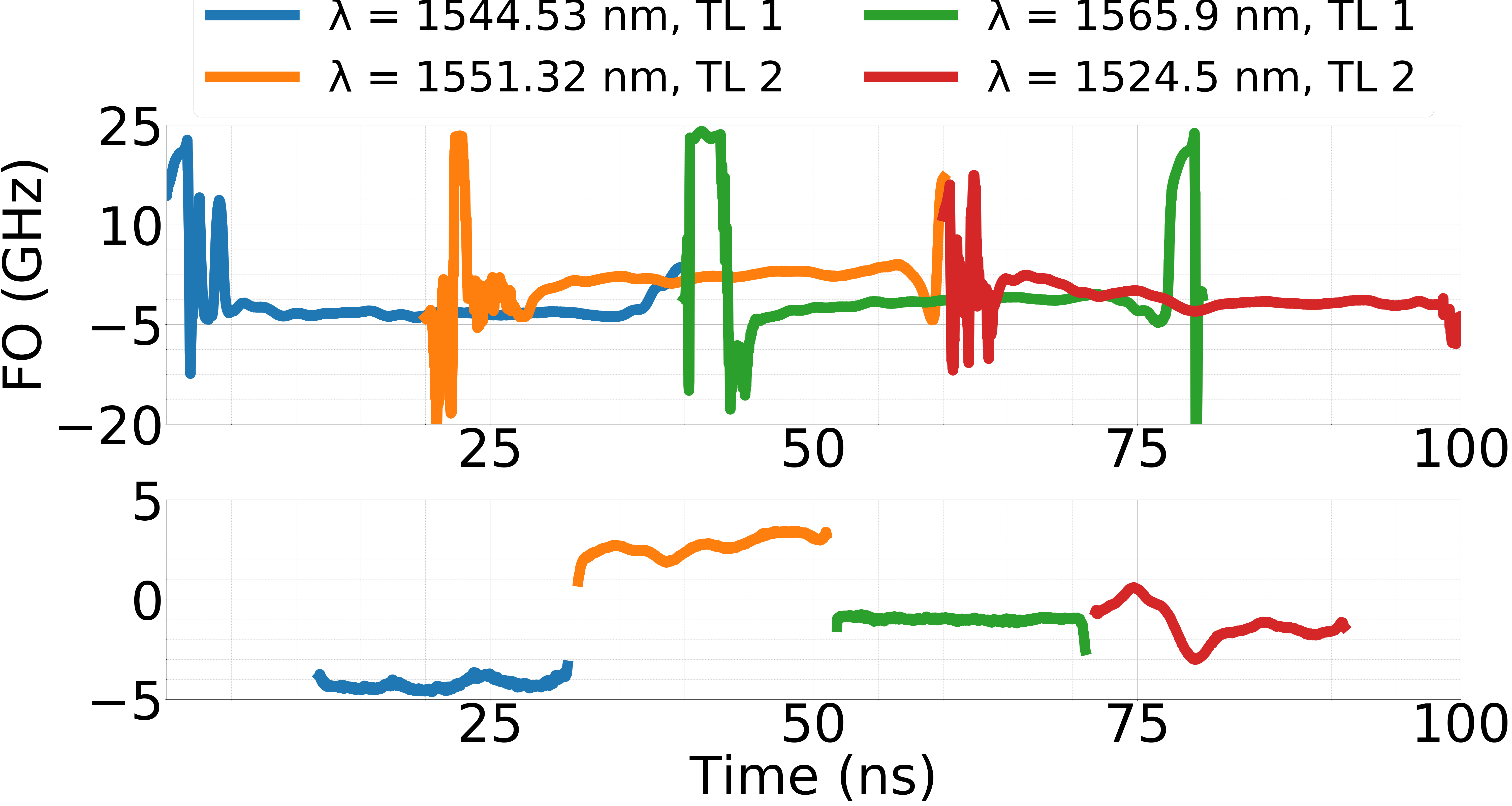}
         \put(-1,45){(f)}
    \end{overpic}
    \\
    \end{tabular}
\vspace{-1em}
\caption{SOA outputs showing (a) step \& (b) PSO rise \& settling times, (c) SWIFT system output, (d) $\lambda$-to-$\lambda$ 90-90\% switching times, (e) optical spectrum of the 21 worst-case channels, (f) frequency offset (FO) of DSDBR (top) \& SWIFT (bottom).}
\vspace{-2em}
\end{figure*}
\vspace{-0.3em}

\vspace{-0.5em}
\subsection{SWIFT module demonstration}
\vspace{-0.3em}
 
After optimisation, the DS-DBR lasers were driven with with 12.5~MHz pre-emphasised square waves, resulting in $\leq$40~ns bursts on each wavelength. The lasers were driven 20~ns out of phase, so that one lased while the other reconfigured. The SOAs were driven by 25~MHz PSO-optimised signals, resulting in 20~ns gates, and aligned to block the first 15~ns and last 5~ns of each laser burst, yielding four wavelength bursts of 20~ns each (see Fig.~2(a)). Fig. 3(c) shows the oscilloscope output for the most difficult switching instance, where DS-DBR laser 1 switched from 1572.48~nm to 1524.11~nm, incurring a large rear current swing of 45~mA. The oscilloscope shows a flat intensity response across each wavelength for 20~ns bursts, thereby providing twice
the granularity reported in \cite{shi2019}. Packet-to-packet power variations are due to slight variations in laser wavelength power; these can be addressed by applying slot specific SOA drive currents (not possible in our setup). Measuring switch time by the 90-90\% transition time, we report switch times for the four transitions of 771, 812, 521, and 792 ps, respectively. These are shown in Fig.~3(d).  
Furthermore, Fig.~3(f) shows the coherent receiver output of the four wavelength slots with and without gating. The observed frequency ripples are a result of the low sample rate of our 250~MS/s AWG that introduce Fourier components to the driving square wave; these can be easily suppressed by using a higher sample rate. Despite this, each slot stays within 5~GHz of its target.  

We repeated this process for each of the channels under test. Fig. 3(e) shows the optical spectrum for all channels, all undergoing gated switching. We measured a worst case value for the side mode suppression ratio of 35~dB, optical power output of 0.8~dBm for a single wavelength (at 1572.48~nm) and corresponding extinction ratio of 22~dB. The fully time-multiplexed optical output power of SWIFT was $>$6~dBm.
 This represents the largest number of sub-ns switching channels from a single sub-system ever reported, supporting 122$\times{}$50~GHz spaced channels. 


\textbf{In conclusion}, we propose a scalable, low power, tuneable wavelength subsystem capable of sub-ns switching. Using pairs of time-multiplexed tuneable lasers, gated by SOAs, we have experimentally demonstrated switching times of less than 900~ps for 122 x 50~GHz channels.
Reliable and fast tuning was achieved for each laser and SOA using regression and particle swarm optimisation AI techniques. This enables automated, device-specific optimisation and represents a critically important technology in OCS architectures, potentially transforming DCN architectures. 
\vspace{-0.2em}
\begin{center}\begin{small}\it{This work is supported by EPSRC (EP/R035342/1), IPES CDT, iCASE and Microsoft Research.}\end{small}
\end{center}
\vspace{-1em}




